\documentclass[11pt]{article}

\usepackage{tikz}
\usepackage{pgf}
\usetikzlibrary{arrows}
\usetikzlibrary{snakes}
\usetikzlibrary{decorations,decorations.pathmorphing,decorations.markings}
\usepackage{fullpage}

\tikzset{graviton/.style={decorate, decoration={snake}, double}}

\usepackage{cancel}
\usepackage{braket}
\usepackage[T1]{fontenc}
\usepackage[utf8]{inputenc}
\usepackage{xcolor}
\usepackage{tikz-feynman}
\usepackage[italian,english]{babel}
\usepackage{url}
\usepackage{siunitx}
\usepackage{amsmath}
\usepackage{amssymb}
\usepackage{graphicx}
\usepackage{graphicx,caption}
\usepackage{siunitx}
\usepackage{bm}
\usepackage{amsfonts}
\usepackage[left=2cm,right=2cm,top=2.5cm,bottom=2.5cm]{geometry}
\usepackage{amsmath}
\usepackage{graphicx}
\usepackage{subfigure}
\usepackage{slashed}
\usepackage{verbatim}
\usepackage[hidelinks]{hyperref}

\let\originaleqref\eqref 
\makeatletter
\renewcommand{\eqref}[1]{%
	\let\ref\@refstar%
	\hyperref[#1]{%%
		~\originaleqref{#1}%
	}%
}
\makeatother

\numberwithin{equation}{section}

\usepackage{tikz}
\usepackage{tikz-feynman}
\usepackage{pgf}
\usetikzlibrary{arrows}
\usetikzlibrary{snakes}
\usetikzlibrary{decorations,decorations.pathmorphing,decorations.markings}

\def\be{\begin{equation}}
\def\ee{\end{equation}}

\usepackage{adjustbox}
\usepackage{standalone}
\usepackage{amsmath}
\newcommand{\beq}{\begin{equation}}
\newcommand{\eeq}{\end{equation}}
\newcommand{\beqa}{\begin{eqnarray}}
\newcommand{\eeqa}{\end{eqnarray}}

\newcommand{\nn}{\nonumber}

\def\id{\leavevmode\hbox{\small1\kern-3.3pt\normalsize1}}

\usepackage{adjustbox}
\usepackage{standalone}
\usepackage{amsmath}

\def\a{\alpha}
\def\b{\beta}

\def\m{\mu}
\def\n{\nu}

\newcommand{\cS}{\mathcal{S}}
\newcommand{\bea}{\begin{eqnarray}}
	\newcommand{\eea}{\end{eqnarray}}

\newcommand{\nc}{\newcommand}
\nc{\rnc}{\renewcommand}
\nc{\D}{\partial}
\nc{\K}{\kappa}
\nc{\bK}{\bar{\K}}
\nc{\bN}{\bar{N}}
\nc{\bq}{\bar{q}}
\nc{\vbq}{\vec{\bar{q}}}
\nc{\g}{\gamma}
\nc{\lrarrow}{\leftrightarrow}
\nc{\rg}{\sqrt{g}}
\nc{\de}{\delta}

\rnc{\(}{\left(}
\rnc{\)}{\right)}
\nc{\q}{\vec{q}}
\nc{\x}{\vec{x}}
\nc{\ep}{\epsilon}
\nc{\tto}{\rightarrow}
\rnc{\inf}{\infty}
\rnc{\Re}{\mathrm{Re}}
\rnc{\Im}{\mathrm{Im}}
\nc{\z}{\zeta}
\nc{\mA}{\mathcal{A}}
\nc{\mB}{\mathcal{B}}
\nc{\mC}{\mathcal{C}}
\nc{\mD}{\mathcal{D}}
\nc{\mN}{\mathcal{N}}
\rnc{\H}{\mathcal{H}}
\rnc{\L}{\mathcal{L}}
\nc{\<}{\langle}
\rnc{\>}{\rangle}
\nc{\fnl}{f_{NL}}
\nc{\gnl}{g_{NL}}
\nc{\fnleq}{f_{NL}^{equil.}}
\nc{\fnlloc}{f_{NL}^{local}}
\nc{\vphi}{\varphi}
\nc{\Lie}{\pounds}
\nc{\half}{\frac{1}{2}}
\nc{\bOmega}{\bar{\Omega}}
\nc{\bLambda}{\bar{\Lambda}}
\nc{\dN}{\delta N}
\nc{\gYM}{g_{\mathrm{YM}}}
\nc{\geff}{g_{\mathrm{eff}}}
\nc{\tr}{\mathrm{tr}}
\nc{\oa}{\stackrel{\leftrightarrow}}
\nc{\IR}{{\rm IR}}
\nc{\UV}{{\rm UV}}

\begin{document}
\vspace{1.5cm}
\begin{center}
{\bf  \Large The Gravitational Chiral Anomaly at Finite Temperature and Density } 
\vspace{0.2cm}
{\bf  \Large }

{\Large \bf  }
\vspace{0.1cm}
{\Large \bf }

 \vspace{0.3cm}
\vspace{1cm}
{\bf $^{(1,2)}$Claudio Corian\`o, $^{(1) (3)}$Mario Cret\`i, $^{(1)}$Stefano Lionetti, $^{(1)}$Riccardo Tommasi \\}

\vspace{1cm}
{\it  $^{(1)}$Dipartimento di Matematica e Fisica, Universit\`{a} del Salento \\
and INFN Sezione di Lecce, Via Arnesano 73100 Lecce, Italy\\
National Center for HPC, Big Data and Quantum Computing\\}
\vspace{0.5cm}
{\it  $^{(2)}$  Institute of Nanotechnology, \\ National Research Council (CNR-NANOTEC), Lecce 73100\\}
\vspace{0.3cm}

{\it $^{(3)}$Center for Biomolecular Nanotechnologies,\\ Istituto Italiano di Tecnologia, Via Barsanti 14,
73010 Arnesano, Lecce, Italy\\}
\vspace{0.5cm}

\begin{abstract}
We investigate the gravitational anomaly vertex $\langle TTJ_5\rangle$ (graviton - graviton - axial current) under conditions of finite density and temperature. Through a direct analysis of perturbative contributions, we demonstrate that neither finite temperature nor finite fermion density affects the gravitational chiral anomaly. These results find application in several contexts, from topological materials to the early universe plasma. They affect the decay of any axion or axion-like particle into gravitational waves, in very dense and hot environments.
\end{abstract}

\end{center} 
\newpage

\section{Introduction} 
The interest on gravitational anomalies has been continuous due to their connections with various theories, including ordinary gauge theories \cite{Kimura:1969iwz,Delbourgo:1972xb}, supergravity \cite{Nielsen:1978ex}, and self-dual antisymmetric fields in string theory \cite{Alvarez-Gaume:1983ihn}. 
The gravitational anomaly $R \tilde{R}$ can manifest in different scenarios, with its implications varying from benign to critical. For example, consider a scenario involving a Dirac fermion interacting with gravity and a vector gauge field. The anomaly, that in this case appears in the divergence of $J_5$, was first computed by Kimura, Delbourgo, and Salam \cite{Kimura:1969iwz,Delbourgo:1972xb}. This specific anomaly poses no threat and holds relevance in phenomenology. \\
Another instance involves a chiral model incorporating a Weyl fermion $\psi_{L/R}$ 
interacting with gravity and a gauge field. In this case, the anomaly emerges in the divergence of $J_{L/R}$, potentially endangering unitarity and renormalization, unless it is canceled \cite{Alvarez-Gaume:1983ihn, Bonora:2023soh}.
In the Standard Model, when $J_5$ is the non-abelian $SU(2)$ gauge current or the hypercharge gauge current, the gravitational anomaly cancel out by summing over the chiral spectrum of each fermion generation. This feature is usually interpreted as an indication of the compatibility of the Standard Model when coupled to a gravitational background, providing an essential constraint on its possible extensions.\\
The correlator $\langle TT J_5\rangle $, where $T$ denotes the stress energy tensor, under examination in this work, has been recently re-investigated  \cite{Coriano:2023gxa} using Conformal Field Theory (CFT) in momentum space methods \cite{Coriano:2013jba,Bzowski:2013sza,Bzowski:2018fql}, previously studied in \cite{Erdmenger:1999xx} using coordinate space methods.\\
Correlators influenced by chiral and conformal anomalies, as well as discrete anomalies, play a vital role in condensed matter theory, particularly in the context of topological materials \cite{Hasan:2010xy,:2013kya,Ambrus:2019khr,Chernodub:2021nff,Arouca:2022psl,Tutschku:2020rjq,Landsteiner:2013sja,Frohlich:2023uqc,Mottola:2023emy,Mottola:2019nui,Landsteiner:2013sja}. The gravitational anomaly has been investigated, in the same context, in other interesting works \cite{Nissinen:2021gke,Nissinen:2018dnq}. Crucial, in this analysis, is the correspondence between thermal stresses and gravity, as summarized by Luttinger's relation \cite{Luttinger:1964zz} connecting a gravitational potential to a thermal gradient \cite{Bermond:2022mjo,Chernodub:2019tsx}.\\
Understanding these phenomena is essential for unraveling the intricate properties of such materials. Since their dynamical contribution in the evolution of topological matter, in the realistic case, is characterised by both thermal effecs and by Fermi surfaces, which break the charge conjugation $C$ invariance of the vacuum, the quantification of such corrections becomes crucial for phenomenology. \\
\subsection{Thermal and density effects}
Considerable attention has been dedicated to examining the effects of finite temperature and density on the axial gauge anomaly in several other contexts \cite{Contreras:1987ku,Reuter:1985zc,Chaturvedi:1985xsm,Das:1987yb,Qian:1994pp,Coriano:2024nhv,Kamada:2022nyt,Kanazawa:2015xna}.
Despite the diverse approaches taken by the various authors to address this issue, a unanimous consensus emerges: such anomaly remains insensitive to corrections from finite temperature and density.\\
In our previous work \cite{Coriano:2024nhv}, we have investigated the general structure of the chiral anomaly vertex $\langle AVV\rangle$, in the presence of chemical potentials in perturbation theory. 
We have classified the minimal number of tensorial structures for the $\langle AVV\rangle$ parameterization and we have provided a direct perturbative identification of their corresponding form factors. Moreover, when the photons are on-shell, we have shown that the entire correlator reduces to the longitudinal anomalous sector.\\
The gravitational axial anomaly has yet to be explored in the context of finite temperature and density. In this paper, we aim to fill this gap by investigating this interaction.
Our approach involves utilizing the real-time Green's function method \cite{Das:1997gg}. Through this technique, we will compute the amplitude of thermal Feynman diagrams. Our objective is to demonstrate that the gravitational axial anomaly remains unchanged, as the contributions from density and temperature cancel out precisely.

\section{Review: Decomposition of the correlator and the perturbative expansion at zero temperature and density}\label{sectiondecompcorr}
In this section, we provide an overview of the general structure for the $\langle TTJ_5\rangle$ correlator in the vacuum, using the longitudinal/transverse-traceless decomposition, commonly referred to as the "$L/T$" decomposition \cite{Bzowski:2013sza}. This allows us to gain insight into its global tensorial structure under zero temperature and density conditions. The decomposition has been used in the case of parity-even anomaly correlators in several previous works involving both 3-point and 4-point functions \cite{Bzowski:2018fql,Coriano:2018bbe,Coriano:2018bsy} \cite{Coriano:2022jkn} and extended to the simplest parity-odd correlator, the $AVV$ in \cite{Coriano:2023hts}. Non conserved vector currents, have been discussed by the same method in \cite{Marotta:2022jrp}. 
The analysis presented here, at zero $T$ and $\mu$, is nonperturbative since it relies on the solution of the conformal constraints by the inclusion of an anomaly pole in order to fix the longitudinal sector of the correlator, and does not require a Lagrangian realization. 
The structure becomes considerably more intricate in the case of finite $T$ and $\mu$, primarily due to the inclusion of a velocity four-vector $\eta$ representing the heat bath. The exhaustive analysis of the vertex entails a tensorial expansion, which is particularly demanding, even in the simpler $AVV$ diagram case \cite{Coriano:2024nhv}.\\
We start by decomposing the energy-momentum tensor $T^{\m \n}$ and the current $J_5^\m$ in terms of their transverse-traceless part and longitudinal ones (also called "local")
\begin{align}
	T^{\mu_i\nu_i}(p_i)&= t^{\mu_i\nu_i}(p_i)+t_{loc}^{\mu_i\nu_i}(p_i),\label{decT}\\
	J_5^{\mu_i}(p_i)&= j_5^{\mu_i}(p_i)+j_{5 \, loc}^{\mu_i}(p_i),\label{decJ}
\end{align}
where
\begin{align}
	\label{loct}
	&t^{\mu_i\nu_i}(p_i)=\Pi^{\mu_i\nu_i}_{\alpha_i\beta_i}(p_i)\,T^{\alpha_i \beta_i}(p_i), &&t_{loc}^{\mu_i\nu_i}(p_i)=\Sigma^{\mu_i\nu_i}_{\alpha_i\beta_i}(p)\,T^{\alpha_i \beta_i}(p_i),\nn\\
	&j_5^{\mu_i}(p_i)=\pi^{\mu_i}_{\alpha_i}(p_i)\,J_5^{\alpha_i }(p_i), \hspace{1ex}&&j_{5\, loc}^{\mu_i}(p_i)=\frac{p_i^{\mu_i}\,p_{i\,\alpha_i}}{p_i^2}\,J_5^{\alpha_i}(p_i),
\end{align}
having introduced the transverse-traceless ($\Pi$), transverse $(\pi)$ and longitudinal ($\Sigma$) projectors, given respectively by 
\begin{align}
	\label{prozero}
	&\pi^{\mu}_{\alpha}  = \delta^{\mu}_{\alpha} - \frac{p^{\mu} p_{\alpha}}{p^2}, \\&
	\Pi^{\mu \nu}_{\alpha \beta}  = \frac{1}{2} \left( \pi^{\mu}_{\alpha} \pi^{\nu}_{\beta} + \pi^{\mu}_{\beta} \pi^{\nu}_{\alpha} \right) - \frac{1}{d - 1} \pi^{\mu \nu}\pi_{\alpha \beta}\label{TTproj}, \\&
	\Sigma^{\mu_i\nu_i}_{\alpha_i\beta_i}=\frac{p_{i\,\beta_i}}{p_i^2}\Big[2\delta^{(\nu_i}_{\alpha_i}p_i^{\mu_i)}-\frac{p_{i\alpha_i}}{(d-1)}\left(\delta^{\mu_i\nu_i}+(d-2)\frac{p_i^{\mu_i}p_i^{\nu_i}}{p_i^2}\right)\Big]+\frac{\pi^{\mu_i\nu_i}(p_i)}{(d-1)}\delta_{\alpha_i\beta_i}\label{Lproj}.
\end{align}
Such decomposition allows to split the vacuum correlation function into the following terms
\begin{equation} \label{eq:splitlongttpart}
	\begin{aligned}
		\left\langle T^{\mu_{1} \n_{1}} T^{\mu_{2} \n_2} J_5^{\mu_{3}}\right\rangle=&\left\langle t^{\mu_{1} \n_{1}} t^{\mu_{2}\n_2} j_5^{\mu_{3}}\right\rangle+\left\langle T^{\mu_{1} \n_{1}} T^{\mu_{2}\n_2} j_{5\, l o c}^{\mu_{3}}\right\rangle+\left\langle T^{\mu_{1} \n_{1}} t_{l o c}^{\mu_{2}\n_2} J_5^{\mu_{3}}\right\rangle+\left\langle t_{l o c}^{\mu_{1} \n_{1}} T^{\mu_{2}\n_2} J_5^{\mu_{3}}\right\rangle \\
		&-\left\langle T^{\mu_{1} \n_{1}} t_{l o c}^{\mu_{2}\n_2} j_{5\, l o c}^{\mu_{3}}\right\rangle-\left\langle t_{l o c}^{\mu_{1} \n_{1}} t_{l o c}^{\mu_{2}\n_2} J_5^{\mu_{3}}\right\rangle-\left\langle t_{l o c}^{\mu_{1} \n_{1}} T^{\mu_{2}\n_2} j_{5\, l o c}^{\mu_{3}}\right\rangle+\left\langle t_{l o c}^{\mu_{1} \n_{1}} t_{l o c}^{\mu_{2}\n_2} j_{5\, l o c}^{\mu_{3}}\right\rangle, 
	\end{aligned}
\end{equation}
where the first contribution on the r.h.s. of the expression above identifies the transverse-traceless part. 
Using the (anomalous) conservation and trace WIs, it is then possible to completely fix all the longitudinal parts, i.e. the terms containing at least one $j_{5 \, loc}^{\mu}$ or $t_{ loc}^{\mu\nu}$. The non-anomalous equations are
\begin{equation}
		\begin{aligned}
			&\delta_{\mu_i \nu_i}\left\langle T^{\mu_1 \nu_1}\left(p_1\right) T^{\mu_2 \nu_2}\left(p_2\right) J_5^{\mu_3}\left(p_3\right)\right\rangle=0,\qquad\qquad && i=\{1,2\} \\&
			p_{i \mu_i}\left\langle T^{\mu_1 \nu_1}\left(p_1\right) T^{\mu_2 \nu_2}\left(p_2\right) J_5^{\mu_3}\left(p_3\right)\right\rangle=0, && i=\{1,2\}
		\end{aligned}
\end{equation} 
Thanks to these WIs, we can eliminate most of terms on the right-hand side of equation \eqref{eq:splitlongttpart}, ending up with only two terms
\begin{equation}
	\left\langle T^{\mu_{1} \n_{1}} T^{\mu_{2} \n_2} J_5^{\mu_{3}}\right\rangle=\left\langle t^{\mu_{1} \n_{1}} t^{\mu_{2}\n_2} j_5^{\mu_{3}}\right\rangle+\left\langle T^{\mu_{1} \n_{1}} T^{\mu_{2}\n_2} j_{5\, l o c}^{\mu_{3}}\right\rangle=\left\langle t^{\mu_{1} \n_{1}} t^{\mu_{2}\n_2} j_5^{\mu_{3}}\right\rangle+\left\langle t^{\mu_{1} \n_{1}} t^{\mu_{2}\n_2} j_{5\, l o c}^{\mu_{3}}\right\rangle.
\end{equation}
The remaining local term in the expression above is then fixed by the anomalous WI of $J_5$
\small
\begin{equation}\label{eq:idwanomlp3}
	p_{3\mu_3}\braket{T^{\mu_1\nu_1}(p_1)T^{\mu_2\nu_2}(p_2)J_5^{\mu_3}(p_3)}= 4\, i \, a_2 \, (p_1 \cdot p_2) \left\{ \left[\varepsilon^{\nu_1 \nu_2 p_1 p_2}\left(g^{\mu_1 \mu_2}- \frac{p_1^{\mu_2} p_2^{\mu_1}}{p_1 \cdot p_2}\right) +\left( \mu_1 \leftrightarrow \nu_1 \right) \right] +\left( \mu_2 \leftrightarrow \nu_2 \right) \right\}.
\end{equation}
\normalsize
where $a_2$ is the anomaly constant. Thanks to the equation above, we can write
\begin{equation} \label{eq:anompolettj}
	\left\langle t^{\mu_{1} \n_{1}} t^{\mu_{2}\n_2} j_{5\, l o c}^{\mu_{3}}\right\rangle= 4 ia_2 \frac{p_3^{\mu_3}}{p_3^2} \, (p_1 \cdot p_2) \left\{ \left[\varepsilon^{\nu_1 \nu_2 p_1 p_2}\left(g^{\mu_1 \mu_2}- \frac{p_1^{\mu_2} p_2^{\mu_1}}{p_1 \cdot p_2}\right) +\left( \mu_1 \leftrightarrow \nu_1 \right) \right] +\left( \mu_2 \leftrightarrow \nu_2 \right) \right\}.
\end{equation}
Notice that in order to identify in field space the effective action associated with this contribution, we need to 
multiply the stress energy tensors by the gravitational fluctuations of the metric $h_{\mu\nu}$  and the axial-vector source $A_\mu$ on the current $J_5$, in the form 

\beq
\mathcal{S}_{anom}\sim a_2 \int d^4 x \, d^4 y \partial \cdot A(x) \Box^{-1}(x,y) R \tilde{R}(y) 
\eeq
In other words, eq$.$ \eqref{eq:anompolettj} can be generated from $\mathcal{S}_{anom}$ by the functional differentiation 
of the latter twice w.r.t. the metric and once w.r.t. $A_\mu$, followed by an ordinary Fourier transform. 
The terms $1/p_3^2$ is the anomaly pole, introduced in order to solve \eqref{eq:idwanomlp3}. \\
Notice that this four-dimensional procedure is quite straightforward in the case of chiral anomaly diagrams, since the conformal Ward identities are exact, i.e. they are not affected by the anomaly. The reason for such behaviour is quite simple, and is due to the fact that there is no renormalzation needed in order to define the correlator at spacetime dimension four. The total anomaly effective action can indeed be decomposed in the form 
\beq
\mathcal{S}_{eff}=\mathcal{S}_{anom} +\mathcal{S}_\perp. 
\eeq
The only remaining term in the reconstruction of the entire correlator is the transverse-traceless part $\left\langle t^{\mu_{1} \n_{1}} t^{\mu_{2}\n_2} j_5^{\mu_{3}}\right\rangle$ related to $\mathcal{S}_\perp$.  
Its explicit form is given by the inclusion of transverse and transverse-traceless projectors $\pi$ and $\Pi$ acting on a tensor structure that is parameterised in terms of a minimal number of independent form factors 
\begin{equation} \label{eq:defX}
	\left\langle t^{\mu_1 \n_1}\left(p_1\right) t^{\mu_2 \n_2}\left(p_2\right) j_5^{\mu_3}\left(p_3\right)\right\rangle=\Pi_{\alpha_1 \beta_1}^{\mu_1 \nu_1}\left(p_1\right) \Pi_{\alpha_2 \b_2}^{\mu_2 \n_2}\left(p_2\right) \pi_{\alpha_3}^{\mu_3}\left(p_3\right) X^{\alpha_1 \beta_1 \alpha_2 \b_2 \alpha_3}
\end{equation}
where $X^{\alpha_1 \beta_1 \alpha_2 \b_2 \alpha_3}$ is a general rank five tensor built by products of metric tensors, momenta and $\varepsilon$ tensors with the appropriate choice of indices. Indeed, as a consequence of the projectors in \eqref{eq:defX}, $X^{\alpha_1 \beta_1 \alpha_2 \b_2 \alpha_3}$ can
not be constructed by using $g_{\a_i \b_i}$, nor by $p_{i \, \a_i}$ with $i =\{1,2,3\}$. 
Using the Schouten identities and imposing the symmetries of the correlators, the general structure of the transverse-traceless part is given by the simplified minimal expression \cite{Coriano:2023gxa}
\begin{equation}\label{eq:decomptt}
	\begin{aligned}
		\langle t^{\mu_{1} \nu_{1}}\left({p}_{1}\right)& t^{\mu_{2} \nu_{2}}\left({p}_{2}\right) j_5^{\mu_{3} }\left({p_3}\right)\rangle=\Pi_{\alpha_{1} \beta_{1}}^{\mu_{1} \nu_{1}}\left({p}_{1}\right) \Pi_{\alpha_{2} \beta_{2}}^{\mu_{2} \nu_{2}}\left({p}_{2}\right) \pi_{\alpha_{3}}^{\mu_{3}}\left({p_3}\right) \bigg[
		\\
		&A_1\varepsilon^{p_1\alpha_1\alpha_2\alpha_3}p_2^{\beta_1}p_3^{\beta_2}
		-A_1(p_1\leftrightarrow p_2) \varepsilon^{p_2\alpha_1\alpha_2\alpha_3}p_2^{\beta_1}p_3^{\beta_2}\\
		&+A_2\varepsilon^{p_1\alpha_1\alpha_2\alpha_3}\delta^{\beta_1\beta_2}-
		A_2(p_1\leftrightarrow p_2)\varepsilon^{p_2\alpha_1\alpha_2\alpha_3}\delta^{\beta_1\beta_2}\\
		&+A_3\varepsilon^{p_1p_2\alpha_1\alpha_2}p_2^{\beta_1}p_3^{\beta_2}p_1^{\alpha_3}
		+A_4\varepsilon^{p_1p_2\alpha_1\alpha_2}\delta^{\beta_1\beta_2}p_1^{\alpha_3}
		\bigg]
	\end{aligned}
\end{equation}
 where $A_3$ and $A_4$ are antisymmetric under the exchange $(p_1\leftrightarrow p_2)$. The expressions of the form factors $A_i$ can be derived in two ways. Either from the solutions of the conformal Ward identities (CWIs), which are expressed in terms of 3K integrals, i.e. parametric integrals of three Bessel functions, or they can be extracted by resorting to the perturbative expansion. 
 The system of equations derived from the CWIs is rather involved, but one can show that the transverse-traceless sector is completely determined in terms of the same coefficient $a_2$ characterising the anomaly constraint. Details can be found in \cite{Coriano:2023gxa}. \\
Both the perturbative and non-perturbative procedures are in complete agreement. In the next section, we will review the former approach.

\subsection{The action and the perturbative realization} 
\begin{figure}[t]  
	\centering
	\includegraphics[scale=0.46]{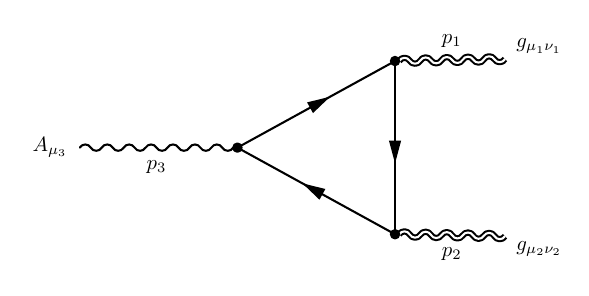}
	\includegraphics[scale=0.46]{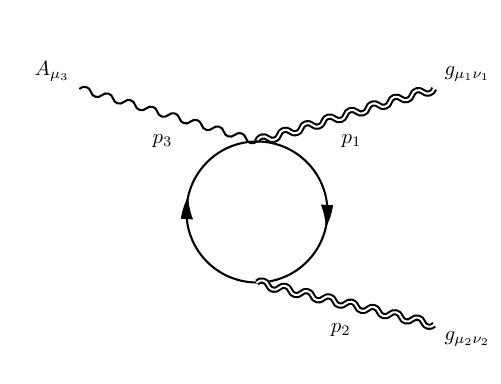}
	\includegraphics[scale=0.46]{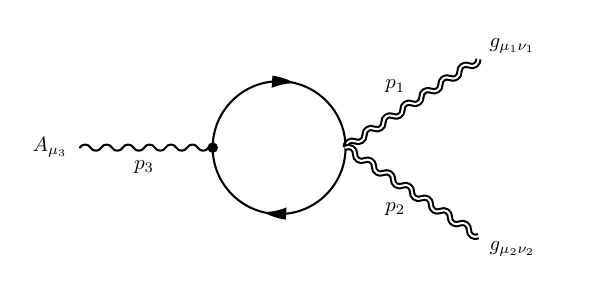}
	\includegraphics[scale=0.46]{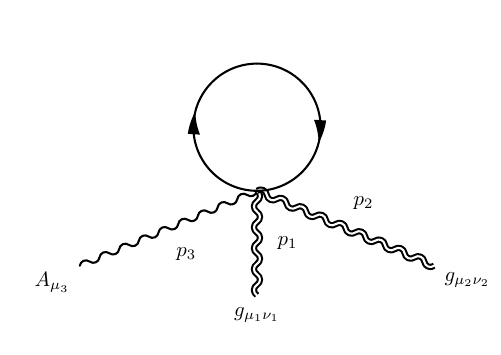}
	\caption{Topologies of Feynman diagrams appearing in the perturbative computation.} 
	\label{fig:feynmdiagr}
\end{figure}

The perturbative evaluation of the correlator at one-loop, can be performed by working in a specific regularization scheme. We define the vacuum partition function 
\be
e^{i \cS[g]} \equiv \int [d \Phi]\, e^{i {{S}_{0}}[\Phi,\, g]}
\ee
with the action given by
\begin{equation}
	S_0=\int d^d x \, \frac{e}{2}\,  e_a^\mu \bigg[{i}  \bar{\psi} \gamma^a\left(D_\mu \psi\right)-{i} \left(D_\mu \bar{\psi}\right) \gamma^a \psi\bigg]
\end{equation}
where $e_a^\mu $ is the vielbein, $e$ is its determinant and $D_\mu$ is the covariant derivative defined as
\begin{equation}
	\begin{aligned}
		& D_\mu \psi=\left(\nabla_\mu+i g \gamma_5 A_\mu\right) \psi=\left(\partial_\mu+i g \gamma_5A_\mu+\frac{1}{2} \omega_{\mu a b} \Sigma^{a b}\right) \psi, \\
		& D_\mu \bar{\psi}=\left(\nabla_\mu-i g \gamma_5 A_\mu\right) \bar{\psi}=\left(\partial_\mu-i g\gamma_5 A_\mu-\frac{1}{2} \omega_{\mu a b} \Sigma^{a b}\right) \bar{\psi}.
	\end{aligned}
\end{equation}
$\Sigma^{ab}$ are the generators of the Lorentz group in the case of a spin 1/2-field, while the spin connection is given by
\begin{equation}
	\omega_{\mu a b} \equiv e_a^\nu\left(\partial_\mu e_{\nu b}-\Gamma_{\mu \nu}^\lambda e_{\lambda b}\right).
\end{equation}
The Latin and Greek indices are related to the (locally) flat basis and the curved background respectively. Using the explicit expression of the generators of the Lorentz group one can re-express the action $S_0$ as follows
\begin{equation}
	S_0=\int d^d x \, e \left[\frac{i}{2} \bar{\psi} e_a^\mu \gamma^a\left(\partial_\mu \psi\right)-\frac{i}{2}\left(\partial_\mu \bar{\psi}\right) e_a^\mu \gamma^a \psi-g A_\mu  \bar{\psi}  e_a^\mu  \gamma^a\gamma_5\psi  +\frac{i}{4} \omega_{\mu a b} e_c^\mu \bar{\psi} \gamma^{a b c} \psi\right]
\end{equation}
with
\begin{equation}
	\gamma^{abc}=\{\Sigma^{ab},\gamma^c \}.
\end{equation}
The explicit expressions of the vertices for such action are reported in Appendix \ref{app:vertices}.
Taking a first variation of the action with respect to the metric one can construct the energy momentum tensor as
\begin{equation}
	T^{\mu \nu}=-\frac{i}{2}\left[\bar{\psi} \gamma^{(\mu} \nabla^{\nu)} \psi-\nabla^{(\mu} \bar{\psi} \gamma^{\nu)} \psi-g^{\mu \nu}\left(\bar{\psi} \gamma^\lambda \nabla_\lambda \psi-\nabla_\lambda \bar{\psi} \gamma^\lambda \psi\right)\right]-g \bar{\psi}\left(g^{\mu \nu} \gamma^\lambda A_\lambda-\gamma^{(\mu} A^{\nu)}\right) \gamma_5 \, \psi,
\end{equation}
We now proceed with the perturbative evaluation.  For this, we use the Breitenlohner-Maison regularization scheme. The topology of the diagrams is shown in Fig. \eqref{fig:feynmdiagr}.
The contribution of the triangle diagrams is given by 
\begin{equation}
	\begin{aligned}
		V_{1}^{\mu_1\nu_1\mu_2\nu_2\mu_3}=-i^3\int \frac{d^d l}{(2\pi)^d}\,\, \frac{\text{tr}\big[V^{\mu_1\nu_1}_{g \bar{\psi}\psi}(l-p_1,l)\, (\cancel{l}-\cancel{p}_1)   V^{\mu_3}_{A \bar{\psi}\psi}
			(\cancel{l}+\cancel{p}_2)
			V^{\mu_2\nu_2}_{g \bar{\psi}\psi}(l,l+p_2)
			\cancel{l}
			\big]}{(l-p_1)^2\,(l+p_2)^2\, l^2} +\textrm{exchange}
	\end{aligned}
\end{equation}
while the bubble diagrams are
\begin{equation}
	\begin{aligned}
		V_2^{\mu_1\nu_1\mu_2\nu_2\mu_3}=-i^2\int \frac{d^d l}{(2\pi)^d}\,\, \frac{\text{tr}\big[V^{\mu_1\nu_1\mu_3}_{gA \bar{\psi}\psi}\, (\cancel{l}+\cancel{p}_2)   V^{\mu_2\nu_2}_{g \bar{\psi}\psi}
			(l,l+p_2) \cancel{l}
			\big]}{(l+p_2)^2\,l^2} +\textrm{exchange}
	\end{aligned}
\end{equation}
and
\begin{equation}
	\begin{aligned}
		V_3^{\mu_1\nu_1\mu_2\nu_2\mu_3}=-i^2\int \frac{d^d l}{(2\pi)^d}\,\, \frac{\text{tr}\big[V^{\mu_1\nu_1\mu_2\nu_2}_{gg \bar{\psi}\psi}(p_1,p_2,l-p_1-p_2,l)\, (\cancel{l}-\cancel{p}_1-\cancel{p}_2)   V^{\mu_3}_{A \bar{\psi}\psi}  \cancel{l}
			\big]}{(l-p_1-p_2)^2\,l^2}. 
	\end{aligned}
\end{equation}
After performing the integration, one can verify  that $V_2^{\mu_1\nu_1\mu_2\nu_2\mu_3}$ vanishes.
Lastly, the tadpole diagram is given by
\begin{equation}
	\begin{aligned}
		V_4^{\mu_1\nu_1\mu_2\nu_2\mu_3}=-i\int \frac{d^d l}{(2\pi)^d}\,\, \frac{\text{tr}\big[V^{\mu_1\nu_1\mu_2\nu_2\mu_3}_{ggA \bar{\psi}\psi}\, \cancel{l}  
			\big]}{l^2}. 
	\end{aligned}
\end{equation}
This last diagram vanishes since it contains the trace of two $\gamma$'s and a $\gamma_5$. The perturbative realization of the correlator will be written down as the sum of all these amplitudes, formally given by the expression
\begin{equation}
	\left\langle T^{\mu_1 \nu_1} T^{\mu_2 \nu_2} J_5^{\mu_3}\right\rangle=4 \sum_{i=1}^4 V_i^{\mu_1 \nu_1 \mu_2 \nu_2 \mu_3}
\end{equation}
The perturbative realization of the correlator satisfies the (anomalous) conservation and trace WIs.  Thus, we can decompose the perturbative correlator as described in previous section in order to match the coefficient of the anomaly $a_2$ with the outcome obtained from the perturbative evaluation. Therefore we reintroduce the two terms of the $L/T$ decomposition, now implemented at the perturbative level 
\begin{equation}
	\left\langle T^{\mu_{1} \n_{1}} T^{\mu_{2} \n_2} J_5^{\mu_{3}}\right\rangle_{pert}=\left\langle t^{\mu_{1} \n_{1}} t^{\mu_{2}\n_2} j_5^{\mu_{3}}\right\rangle_{pert}+\left\langle t^{\mu_{1} \n_{1}} t^{\mu_{2}\n_2} j_{5\, l o c}^{\mu_{3}}\right\rangle_{pert}.
\end{equation}
In particular, the anomalous pole is found to be
\begin{equation}
	\left\langle t^{\mu_{1} \n_{1}} t^{\mu_{2}\n_2} j_{5\, l o c}^{\mu_{3}}\right\rangle_{pert}= \frac{g}{96\pi^2}\,  \frac{p_3^{\mu_3}}{p_3^2} \, (p_1 \cdot p_2) \left\{ \left[\varepsilon^{\nu_1 \nu_2 p_1 p_2}\left(g^{\mu_1 \mu_2}- \frac{p_1^{\mu_2} p_2^{\mu_1}}{p_1 \cdot p_2}\right) +\left( \mu_1 \leftrightarrow \nu_1 \right) \right] +\left( \mu_2 \leftrightarrow \nu_2 \right) \right\},
\end{equation}
which corresponds to eq$.$ \eqref{eq:anompolettj} when we set 
\begin{equation} \label{eq:coeffperturbativo}
	a_2= - \frac{i g}{384 \pi^2} \, .
\end{equation}
The transverse-traceless part $\left\langle t^{\mu_{1} \n_{1}} t^{\mu_{2}\n_2} j_5^{\mu_{3}}\right\rangle$ can be expressed in terms of four form factors as described in eq$.$ \eqref{eq:decomptt}. The perturbative calculation in four dimensions gives
\begin{equation} \label{eq:pertresultff}
	\begin{aligned}
		A_1&=\frac{g p_2^2}{24 \pi^2 \lambda^4}
		\Bigg\{A_{11} 
		+A_{12} \log\left(\frac{p_1^2}{p_2^2}\right)+A_{13} \log\left(\frac{p_1^2}{p_3^2}\right)
		+A_{14} \, \,C_0(p_1^2,p_2^2,p_3^2)
		\Bigg\}, \\[8pt]
		A_2&=\frac{gp_2^2}{48\pi^2\lambda^3}\Bigg\{
		A_{21}
		+A_{22} \log\left(\frac{p_1^2}{p_2^2}\right)+A_{23} \log\left(\frac{p_1^2}{p_3^2}\right)
		+A_{24} \, \, C_0(p_1^2,p_2^2,p_3^2)
		\Bigg\}, \\[8pt]
		A_3&=0,\\[8pt]
		A_4&=0\\[8pt],
	\end{aligned}
\end{equation}
where $C_0$ in Minkowski space is the master integral 
\begin{equation}
	C_{0}(p_1^2,p_2^2,p_3^2) \equiv \frac{1}{ i \pi^2}  \int d^d l \frac{1}{l^2(l-p_1)^2(l+p_2)^2}
\end{equation}
and we have introduced the functions $A_{ij}$, reported in the Appendix \ref{app:defAij}.
At this stage, after performing the match of the pole contribution with the perturbative expansion, we proceed by decorating the diagrams with the new propagators at finite density and temperature.

\section{The perturbative realization at finite temperature and density}
In order to compute the finite density and temperature corrections to the gravitational anomaly we need to consider the same Feynman diagrams defined in the vacuum case. However, now we have to replace the usual fermionic propagator with its generalization in a hot medium in the real time formulation. We use the expression 
\begin{equation}\label{eq:propmassivtemp}
	\begin{aligned}
		S_F\equiv (\not k+m) \,G_F=(\not k+m)\Bigg\{\frac{1}{ k^2-m^2}+2 \pi i  \delta\left(k^2-m^2\right)\left[\frac{\theta\left(k_0\right)}{e^{\beta(E-\mu)}+1}+\frac{\theta\left(-k_0\right)}{e^{\beta(E+\mu)}+1}\right]\Bigg\}\,
	\end{aligned}
\end{equation}
with the fermion mass $m$ set to zero. Such expression can also be formulated covariantly by introducing the four-vector $\eta$, defining the velocity of the heath-bath.\\
The contribution of the triangle diagrams is now given by 
\small
\begin{equation}
	\begin{aligned}
		&V_{1}^{\mu_1\nu_1\mu_2\nu_2\mu_3}=\\&-i^3\int \frac{d^d l}{(2\pi)^d}\,\, \text{tr}\bigg[V^{\mu_1\nu_1}_{g \bar{\psi}\psi}(l-p_1,l)\, (\cancel{l}-\cancel{p}_1)   V^{\mu_3}_{A \bar{\psi}\psi}
		(\cancel{l}+\cancel{p}_2)
		V^{\mu_2\nu_2}_{g \bar{\psi}\psi}(l,l+p_2)
		\cancel{l}
		\bigg]G_F(l-p_1)G_F(l+p_2)G_F(l)+\textrm{exchange}
	\end{aligned}
\end{equation}
\normalsize
while the bubble diagrams are
\begin{equation}
	\begin{aligned}
		V_2^{\mu_1\nu_1\mu_2\nu_2\mu_3}=-i^2\int \frac{d^d l}{(2\pi)^d}\,\, \text{tr}\bigg[V^{\mu_1\nu_1\mu_3}_{gA \bar{\psi}\psi}\, (\cancel{l}+\cancel{p}_2)   V^{\mu_2\nu_2}_{g \bar{\psi}\psi}
		(l,l+p_2) \cancel{l}
		\bigg] G_F(l+p_2)G_F(l)+\textrm{exchange}
	\end{aligned}
\end{equation}
and
\begin{equation}
	\begin{aligned}
		V_3^{\mu_1\nu_1\mu_2\nu_2\mu_3}=-i^2\int \frac{d^d l}{(2\pi)^d}\,\, \text{tr}\bigg[V^{\mu_1\nu_1\mu_2\nu_2}_{gg \bar{\psi}\psi}(p_1,p_2,l-p_1-p_2,l)\, (\cancel{l}-\cancel{p}_1-\cancel{p}_2)   V^{\mu_3}_{A \bar{\psi}\psi}  \cancel{l}
		\bigg]G_F(l-p_1-p_2)G_F(l). 
	\end{aligned}
\end{equation}
As previously mentioned, in the case of zero temperature and density, $V_2$ vanishes. The proof relies on Lorentz symmetry, which is violated by temperature and density effects. Consequently, $V_2$ now exhibits non-vanishing thermal contributions that cannot be discarded. Furthermore, the presence of $V_2$ is essential to demonstrate the cancellation of all diagram contributions to the gravitational anomaly. Lastly, similar to the scenario of zero temperature and density, the tadpole diagram vanishes as it contains the trace of two $\gamma$'s and a $\gamma_5$.\\
We can now decompose $G_F$ into a standard contribution to the Fermi propagator $G_0$ and the finite density and temperature corrections  $G_1$ 
\begin{equation}
	\begin{aligned}
		G_F=G_0+G_1,\qquad\qquad G_0=\frac{1}{ k^2-m^2},\qquad\qquad
		G_1=2 \pi i  \delta\left(k^2-m^2\right)\left[\frac{\theta\left(k_0\right)}{e^{\beta(E-\mu)}+1}+\frac{\theta\left(-k_0\right)}{e^{\beta(E+\mu)}+1}\right]
	\end{aligned}
\end{equation}
Then, we can split the $\langle TTJ_5\rangle$ correlator into four different parts depending on the number of $G_1$ contained in the loop integrals
\begin{equation}\label{eq:decompnprop}
	\begin{aligned}
		\left\langle T T J_5\right\rangle=\left\langle T TJ_5\right\rangle^{(0)}+
		\left\langle T TJ_5\right\rangle^{(1)}+
		\left\langle T T J_5\right\rangle^{(2)}+
		\left\langle T T J_5\right\rangle^{(3)}
	\end{aligned}
\end{equation}
$\left\langle T TJ_5\right\rangle^{(0)}$ represents the zero density and temperature part that was computed in the previous section, $\left\langle T TJ_5\right\rangle^{(1)}$ contains only one $G_1$ and so on. Note that the triangles diagrams contribute to all the four terms on the right-hand side of eq$.$ \eqref{eq:decompnprop}, while the bubble diagrams to not contribute to $\left\langle T TJ_5\right\rangle^{(3)}$ since they contain two fermionic propagators $G_F$.

\section{The gravitational anomaly at finite temperature and density}
In this section, we examine the longitudinal anomalous sector of $J_5$, showing that it is protected from finite density and temperature effects.
To achieve this, we dissect the terms associated with finite density and temperature corrections in eq. \eqref{eq:decompnprop} separately. We will illustrate that each of these terms vanishes upon contraction with the momentum of the axial current, $p_3^{\mu_3}$.
The sole surviving term is the zero density and temperature one which showcases the effect of the gravitational anomaly
\begin{equation} \label{eq:anomnoncorrettatFandD}
	\begin{aligned}
			p_{3\,\mu_3} &\left\langle T^{\mu_{1} \n_{1}} T^{\mu_{2} \n_2} J_5^{\mu_{3}}\right\rangle=	p_{3\,\mu_3}\left\langle T^{\mu_{1} \n_{1}} T^{\mu_{2} \n_2} J_5^{\mu_{3}}\right\rangle^{(0)}=\\&\qquad\qquad
			= 4\, i \, a_2 \, (p_1 \cdot p_2) \left\{ \left[\varepsilon^{\nu_1 \nu_2 p_1 p_2}\left(g^{\mu_1 \mu_2}- \frac{p_1^{\mu_2} p_2^{\mu_1}}{p_1 \cdot p_2}\right) +\left( \mu_1 \leftrightarrow \nu_1 \right) \right] +\left( \mu_2 \leftrightarrow \nu_2 \right) \right\}.
	\end{aligned}
\end{equation}
The proof of such statement may appear miraculous, as exceedingly long expressions cancel out. Furthermore, in order to achieve this, we do not need to specify the explicit form of $G_1$, except for the fact that it contains a Dirac delta. There's also no requirement to perform the loop integral since the cancellations occur within the integrand itself.\\ 
This procedure has been previously applied to the more simple case of the gauge chiral anomaly \cite{Contreras:1987ku,Qian:1994pp}.
As we will see, in the case of the gravitational chiral anomaly, the computations are significantly lengthier and require the use of Schouten identities, which relate different tensor structures.
We now proceed with our proof.

\subsection{$\mathbf{\left\langle T TJ_5\right\rangle^{(1)}}$}
\begin{figure}
\centering
\begin{tikzpicture}[scale=0.7]
	\clip (0,0) circle (2) ;
	\draw[postaction={decorate, decoration={
			markings,
			mark=at position 1/6 with {\draw (-4pt, -4pt) -- (4pt, 4pt); } } } ] (180:1) -- (60:1) -- 	(-60:1) --	cycle;
	\draw[snake=coil, segment aspect=0] (180:1) -- +(-2,0);
	\draw[graviton] (60:1) -- +(2,0);
	\draw[graviton] (-60:1) -- +(2,0);
	\vspace{2cm}
\end{tikzpicture} 
\begin{tikzpicture}[scale=0.7]
	\clip (0,0) circle (2) ;
	\draw[postaction={decorate, decoration={
			markings,
			mark=at position 3/6 with {\draw (-4pt, -4pt) -- (4pt, 4pt); } } } ] (180:1) -- (60:1) -- 	(-60:1) --	cycle;
	\draw[snake=coil, segment aspect=0] (180:1) -- +(-2,0);
	\draw[graviton] (60:1) -- +(2,0);
	\draw[graviton] (-60:1) -- +(2,0);
	\vspace{2cm}
\end{tikzpicture} 
\begin{tikzpicture}[scale=0.7]
	\clip (0,0) circle (2) ;
	\draw[postaction={decorate, decoration={
			markings,
			mark=at position 5/6 with {\draw (-4pt, -4pt) -- (4pt, 4pt); } } } ] (180:1) -- (60:1) -- 	(-60:1) --	cycle;
	\draw[snake=coil, segment aspect=0] (180:1) -- +(-2,0);
	\draw[graviton] (60:1) -- +(2,0);
	\draw[graviton] (-60:1) -- +(2,0);
	\vspace{2cm}
\end{tikzpicture} 
\\
\begin{tikzpicture}[scale=0.7]
	\clip (0,0) circle (2) ;
	\draw[postaction={decorate, decoration={
			markings,
			mark=at position 0 with {\draw (-4pt, -4pt) -- (4pt, 4pt); } } } ] (0,0) circle (1);
	\draw[snake=coil, segment aspect=0] (90:1) -- +(-2,.5);
	\draw[graviton, segment aspect=0] (90:1) -- +(2,.5);
	\draw[graviton, segment aspect=0] (-90:1) -- +(2,-.5);
	\vspace{2cm}
\end{tikzpicture}
\begin{tikzpicture}[scale=0.7]
	\clip (0,0) circle (2) ;
	\draw[postaction={decorate, decoration={
			markings,
			mark=at position 1/2 with {\draw (-4pt, -4pt) -- (4pt, 4pt); } } } ] (0,0) circle (1);
	\draw[snake=coil, segment aspect=0] (90:1) -- +(-2,.5);
	\draw[graviton, segment aspect=0] (90:1) -- +(2,.5);
	\draw[graviton, segment aspect=0] (-90:1) -- +(2,-.5);
	\vspace{2cm}
\end{tikzpicture}
\begin{tikzpicture}[scale=0.6]
	\clip (0,0) circle (2.5) ;
	\draw[postaction={decorate, decoration={
			markings,
			mark=at position 1/4 with {\draw (-4pt, -4pt) -- (4pt, 4pt); } } } ]  (0,0) circle (1);
	\draw[snake=coil, segment aspect=0] (180:1) -- +(-2,0);
	\draw[graviton, segment aspect=0] (0:1) -- +(2,.5);
	\draw[graviton, segment aspect=0] (0:1) -- +(2,-.5);
	\vspace{2cm}
\end{tikzpicture}
\begin{tikzpicture}[scale=0.6]
	\clip (0,0) circle (2.5) ;
	\draw[postaction={decorate, decoration={
			markings,
			mark=at position 3/4 with {\draw (-4pt, -4pt) -- (4pt, 4pt); } } } ]  (0,0) circle (1);
	\draw[snake=coil, segment aspect=0] (180:1) -- +(-2,0);
	\draw[graviton, segment aspect=0] (0:1) -- +(2,.5);
	\draw[graviton, segment aspect=0] (0:1) -- +(2,-.5);
	\vspace{2cm}
\end{tikzpicture}
\caption{Topologies of diagrams contributing to ${\left\langle T TJ_5\right\rangle^{(1)}}$. The bar on the fermions' lines denotes the insertion of a hot propagator $G_1$}
\end{figure}
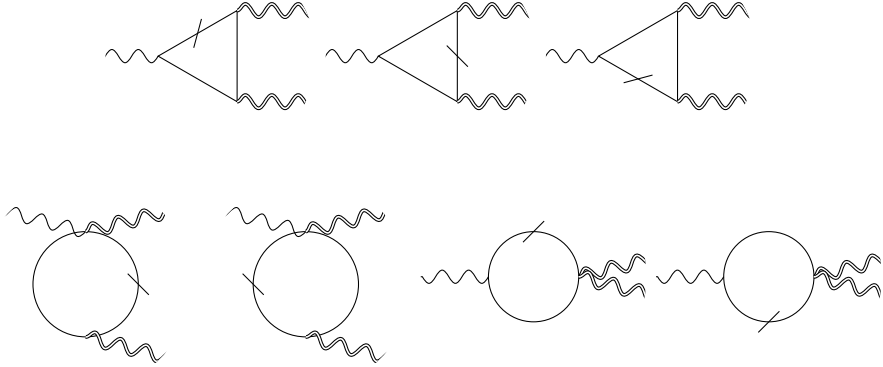
We start by considering $\left\langle T TJ_5\right\rangle^{(1)}$. 
The terms contributing to $\left\langle T TJ_5\right\rangle^{(1)}$ contain only one thermal correction $G_1$ but with different momenta as arguments
\begin{equation}
	\begin{gathered}
		G_1(l),\qquad G_1(l+p_1),\qquad G_1(l-p_1),\qquad G_1(l+p_2),\qquad G_1(l-p_2),\\G_1(l+p_3),\qquad G_1(l-p_3),\qquad G_1(l-p_1+p_2),\qquad G_1(l+p_1-p_2) 
	\end{gathered}
\end{equation}
However, since we are dealing with finite integrals, we can perform multiple shifts in the loop momentum to ensure that the terms inside ${\left\langle T TJ_5\right\rangle^{(1)}}$ only have one single dependence, for example $G_1(l)$. We can then write the contracted correlator in the following form
\begin{equation}
	\begin{aligned}
		p_{3\,\mu_3}  \left\langle T^{\mu_{1} \n_{1}} T^{\mu_{2} \n_2} J_5^{\mu_{3}}\right\rangle^{(1)}&=\int \frac{d^d l}{(2\pi)^d} \, W_{1}^{\mu_1\nu_1\mu_2\nu_2}(p_1,p_2,l)\, G_1(l)
	\end{aligned}
\end{equation}
The dependence on the temperature and chemical potential in the integral above is contained only in $G_1(l)$. 
$W_{1}^{\mu_1\nu_1\mu_2\nu_2}$ is a parity-odd tensor constructed with the momenta $p_1$ $p_2$ and $l$. It is symmetric under the exchange $\{ \mu_1 \leftrightarrow \nu_1\}$ and $\{ \mu_2 \leftrightarrow \nu_2\}$ and $\{ (\mu_1 ,\nu_1, p_1) \leftrightarrow (\mu_2 ,\nu_2, p_2)  \}$.
The explicit initial expression for $W_{1}^{\mu_1\nu_1\mu_2\nu_2}$ is extremely long but one can significantly simplify it by utilizing a set of tensorial relations, known as Schouten identities, which are detailed in Appendix \ref{app:schout}. These identities arise from the dimensional degeneracies of tensor structures, given that we are working in $d=4$.\\
Surprisingly, by applying all the Schouten identities reported in the Appendix \ref{app:schout} and then setting $l^2=0$ due to the $\delta(l^2)$ contained in $G_1(l)$, one is able to prove that 
\begin{equation}
	0=W_1^{\mu_1\nu_1\mu_2\nu_2}\big|_{l^2=0}.
\end{equation}
Therefore
\begin{equation}
	\begin{aligned}
		p_{3\,\mu_3}  \left\langle T^{\mu_{1} \n_{1}} T^{\mu_{2} \n_2} J_5^{\mu_{3}}\right\rangle^{(1)}&=0
	\end{aligned}
	\end{equation}
which means that the $\langle TTJ_5 \rangle^{(1)}$ contributions do not modify the axial anomalous Ward identity.

\subsection{$\mathbf{\left\langle T TJ_5\right\rangle^{(2)} }$}
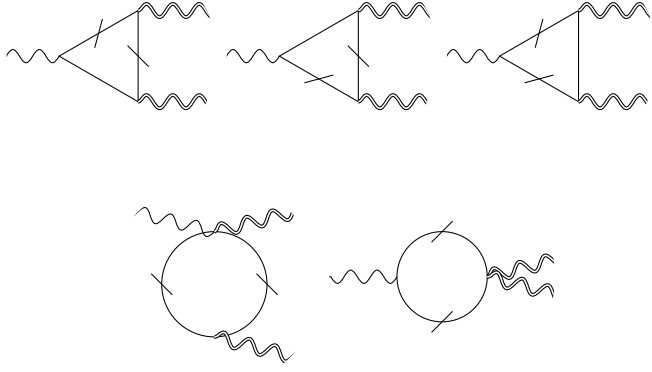
\begin{figure}
\centering
\begin{tikzpicture}[scale=0.7]
	\clip (0,0) circle (2) ;
	\draw[postaction={decorate, decoration={
			markings,
			mark=between positions 1/6 and 3/6 step 1/3 with {\draw (-4pt, -4pt) -- (4pt, 4pt); } } } ] (180:1) -- (60:1) -- 	(-60:1) --	cycle;
	\draw[snake=coil, segment aspect=0] (180:1) -- +(-2,0);
	\draw[graviton] (60:1) -- +(2,0);
	\draw[graviton] (-60:1) -- +(2,0);
	\vspace{2cm}
\end{tikzpicture} 
\begin{tikzpicture}[scale=0.7]
	\clip (0,0) circle (2) ;
	\draw[postaction={decorate, decoration={
			markings,
			mark=between positions 3/6 and 5/6 step 1/3 with {\draw (-4pt, -4pt) -- (4pt, 4pt); } } } ] (180:1) -- (60:1) -- 	(-60:1) --	cycle;
	\draw[snake=coil, segment aspect=0] (180:1) -- +(-2,0);
	\draw[graviton] (60:1) -- +(2,0);
	\draw[graviton] (-60:1) -- +(2,0);
	\vspace{2cm}
\end{tikzpicture} 
\begin{tikzpicture}[scale=0.7]
	\clip (0,0) circle (2) ;
	\draw[rotate=120, postaction={decorate, decoration={
			markings,
			mark=between positions 1/6 and 3/6 step 1/3 with {\draw (-4pt, -4pt) -- (4pt, 4pt); } } } ] (180:1) -- (60:1) -- 	(-60:1) --	cycle;
	\draw[snake=coil, segment aspect=0] (180:1) -- +(-2,0);
	\draw[graviton] (60:1) -- +(2,0);
	\draw[graviton] (-60:1) -- +(2,0);
	\vspace{2cm}
\end{tikzpicture} 
\\
\begin{tikzpicture}[scale=0.7]
	\clip (0,0) circle (2) ;
	\draw[postaction={decorate, decoration={
			markings,
			mark=between positions 0 and 1/2 step 1/2 with {\draw (-4pt, -4pt) -- (4pt, 4pt); } } } ] (0,0) circle (1);
	\draw[snake=coil, segment aspect=0] (90:1) -- +(-2,.5);
	\draw[graviton, segment aspect=0] (90:1) -- +(2,.5);
	\draw[graviton, segment aspect=0] (-90:1) -- +(2,-.5);
	\vspace{2cm}
\end{tikzpicture}
\begin{tikzpicture}[scale=0.6]
	\clip (0,0) circle (2.5) ;
	\draw[postaction={decorate, decoration={
			markings,
			mark=between positions 1/4 and 3/4 step 1/2 with {\draw (-4pt, -4pt) -- (4pt, 4pt); } } } ]  (0,0) circle (1);
	\draw[snake=coil, segment aspect=0] (180:1) -- +(-2,0);
	\draw[graviton, segment aspect=0] (0:1) -- +(2,.5);
	\draw[graviton, segment aspect=0] (0:1) -- +(2,-.5);
	\vspace{2cm}
\end{tikzpicture}
\caption{Topologies of diagrams contributing to ${\left\langle T TJ_5\right\rangle^{(2)}}$. They are characterized by the insertion of two hot propagators $G_1$}
\end{figure}
The procedure for $\left\langle T TJ_5\right\rangle^{(2)}$ is very similar to the one followed in the previous subsection. Considering our parametrization of the loop integral, the combinations in which the thermal corrections $G_1$ appear in $\left\langle T TJ_5\right\rangle^{(2)}$ are
\begin{equation}
	\begin{gathered}
		G_1(l)G_1(l+p_1),\qquad G_1(l)G_1(l+p_2),\qquad G_1(l)G_1(l+p_3),\\ G_1(l)G_1(l-p_2),\qquad G_1(l)G_1(l-p_1),\qquad G_1(l+p_1)G_1(l-p_2),\qquad G_1(l-p_1)G_1(l+p_2).
	\end{gathered}
\end{equation}
Since the integrals are finite, we can perform multiple shifts to the loop momentum, reducing all the combinations above to only three terms
\begin{equation}\label{eq:qwerl}
	\begin{aligned}
		p_{3\,\mu_3}  \left\langle T^{\mu_{1} \n_{1}} T^{\mu_{2} \n_2} J_5^{\mu_{3}}\right\rangle^{(2)}=&\int \frac{d^d l}{(2\pi)^d} \, W_{2,1}^{\mu_1\nu_1\mu_2\nu_2}(p_1,p_2,l)\, G_1(l) G_1(l+p_1)\\&+W_{2,2}^{\mu_1\nu_1\mu_2\nu_2}(p_1,p_2,l)\, G_1(l) G_1(l+p_2)+W_{2,3}^{\mu_1\nu_1\mu_2\nu_2}(p_1,p_2,l)\, G_1(l) G_1(l+p_3)
	\end{aligned}
\end{equation}
The dependence on the temperature and chemical potential in the integrals above is contained in $G_1$. 
The tensors $W_{2,\,i}^{\mu_1\nu_1\mu_2\nu_2}$ are constructed with the momenta $p_1$ $p_2$ and $l$. They are parity-odd and symmetric under the exchange $\{ \mu_1 \leftrightarrow \nu_1\}$ and $\{ \mu_2 \leftrightarrow \nu_2\}$. Moreover, due to the Bose symmetry, we can write
\begin{equation}
	W_{2,1}^{\mu_1\nu_1\mu_2\nu_2}(p_1,p_2,l)=W_{2,2}^{\mu_2\nu_2\mu_1\nu_1}(p_2,p_1,l),\qquad\qquad W_{2,3}^{\mu_1\nu_1\mu_2\nu_2}(p_1,p_2,l)=W_{2,3}^{\mu_2\nu_2\mu_1\nu_1}(p_2,p_1,l)
\end{equation}
The explicit initial expression for $W_{2,i}^{\mu_1\nu_1\mu_2\nu_2}$ is extremely long but one can significantly simplify it by utilizing the Schouten identities. 
Indeed, by applying all the identities reported in the Appendix \ref{app:schout} and using the fact that $G_1$ contains delta functions, we can prove that all three terms in eq$.$ \eqref{eq:qwerl} vanish individually
\begin{equation}
	0=W_{2,1}\big|_{l^2=(l+p_1)^2=0},\qquad\qquad	0=W_{2,2}\big|_{l^2= (l+p_2)^2=0},\qquad\qquad 0= W_{2,3}\big\rvert_{l^2= (l+p_3)^2=0}.
\end{equation}
 Therefore, we have
\begin{equation}
	\begin{aligned}
		p_{3\,\mu_3}  \left\langle T^{\mu_{1} \n_{1}} T^{\mu_{2} \n_2} J_5^{\mu_{3}}\right\rangle^{(2)}&=0
	\end{aligned}
\end{equation}
which means that the $\langle TTJ_5 \rangle^{(2)}$ contributions do not modify the axial anomalous Ward identity.

\subsection{$\mathbf{\left\langle T TJ_5\right\rangle^{(3)}}$}
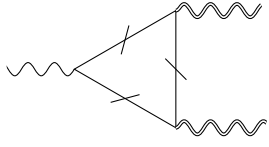
\begin{figure}
\centering
\begin{tikzpicture}[scale=0.9]
	\clip (0,0) circle (2) ;
	\draw[postaction={decorate, decoration={
			markings,
			mark=between positions 1/6 and 5/6 step 1/3 with {\draw (-4pt, -4pt) -- (4pt, 4pt); } } } ] (180:1) -- (60:1) -- 	(-60:1) --	cycle;
	\draw[snake=coil, segment aspect=0] (180:1) -- +(-2,0);
	\draw[graviton] (60:1) -- +(2,0);
	\draw[graviton] (-60:1) -- +(2,0);
	\vspace{2cm}
\end{tikzpicture} 
\caption{Topology of diagram contributing to ${\left\langle T TJ_5\right\rangle^{(3)}}$. Such diagram is characterized by the insertion of three hot propagators $G_1$}
\end{figure}
Only triangle diagrams contribute to ${\left\langle T TJ_5\right\rangle^{(3)}}$ since the bubble diagrams have only two fermionic propagators. The combination of momenta that appear as arguments of $G_1$ are
\begin{equation}
	G_1(l)G_1(l-p_1)G_1(l+p_2),\qquad \qquad G_1(l)G_1(l+p_1)G_1(l-p_2).
\end{equation}
Such combination of $G_1$ can not be further reduced by shift in the loop momentum as in the previous case. Therefore, we can write
\begin{equation}\label{eq:qwerl}
	\begin{aligned}
		p_{3\,\mu_3}  \left\langle T^{\mu_{1} \n_{1}} T^{\mu_{2} \n_2} J_5^{\mu_{3}}\right\rangle^{(3)}=&\int \frac{d^d l}{(2\pi)^d} \, W_{3,1}^{\mu_1\nu_1\mu_2\nu_2}(p_1,p_2,l)\, G_1(l)G_1(l-p_1)G_1(l+p_2) \\&\qquad\qquad +W_{3,2}^{\mu_1\nu_1\mu_2\nu_2}(p_1,p_2,l)\,G_1(l)G_1(l+p_1)G_1(l-p_2)
	\end{aligned}
\end{equation}
$W_{3,\,i}^{\mu_1\nu_1\mu_2\nu_2}$ are parity-odd tensors that depend on the momenta $p_1$ $p_2$ and $l$. They are symmetric under the exchange $\{ \mu_1 \leftrightarrow \nu_1\}$ and $\{ \mu_2 \leftrightarrow \nu_2\}$. Moreover, they are related to each other due to the Bose symmetry
\begin{equation}
	W_{3,1}^{\mu_1\nu_1\mu_2\nu_2}(p_1,p_2,l)=W_{3,2}^{\mu_2\nu_2\mu_1\nu_1}(p_2,p_1,l),
\end{equation}
Once again, we can use the Schouten identities and the fact that $G_1$ contains delta functions in order to prove that 
\begin{equation}
	0=W_{3,1}\big|_{l^2=(l-p_1)^2=(l+p_2)^2=0},\qquad\qquad	0=W_{3,2}\big|_{l^2=(l+p_1)^2=(l-p_2)^2=0}
\end{equation}
Therefore, we have
\begin{equation}
	\begin{aligned}
		p_{3\,\mu_3}  \left\langle T^{\mu_{1} \n_{1}} T^{\mu_{2} \n_2} J_5^{\mu_{3}}\right\rangle^{(3)}&=0
	\end{aligned}
\end{equation}
This completes the proof that the longitudinal axial WI is not modified w.r.t. the vacuum part and the solution of the longitudinal equation is still given by the exchange of an anomaly pole, as shown in \eqref{eq:anomnoncorrettatFandD}. 

\section{Conclusions}
In this paper, we have examined the gravitational anomaly vertex $\langle TTJ_5\rangle$ in a hot and dense medium. We have shown that the anomaly is protected from finite density and temperature corrections. \\
It would be interesting to investigate how dilatations and special conformal transformations are broken in this context \cite{Marchetto:2023fcw}. \\
Our result has application to the decay of an axion or any axion-like particle into gravitational waves, as well as in the production of chiral currents from gravitational waves, in very dense and hot environments. Furthermore, the protection of the gravitational axial anomaly against finite density and temperature corrections presents intriguing experimental opportunities within condensed matter theory, particularly in the context of topological materials. We hope to return to the investigation of these topics in another work.
\\
\\
\centerline{\bf Acknowledgements}
This work is partially funded by the European Union, Next Generation EU, PNRR project "National Centre for HPC, Big Data and Quantum Computing", project code CN00000013; by INFN, inziativa specifica {\em QG-sky} and by the the grant PRIN 2022BP52A MUR "The Holographic Universe for all Lambdas" Lecce-Naples. 

\appendix

\section{Vertices} \label{app:vertices}
In this section we list the explicit expression of all the vertices needed for the perturbative analysis of the $\langle TTJ_5\rangle$ correlator. The momenta of the gravitons and the axial boson are all incoming as well as the momentum indicated with $k_1$. The momentum $k_2$ instead is outgoing. In order to simplify the notation, we introduce the tensor components
\begin{equation}
	\begin{aligned}
		&A^{\mu\nu\rho\sigma}\equiv g^{\mu  \nu } g^{\rho  \sigma }-\frac{1}{2} \left(g^{\mu  \rho } g^{\nu  \sigma }+g^{\mu  \sigma } g^{\nu  \rho }\right)
		\\&
		B^{\mu\nu\rho\sigma\alpha\beta}\equiv g^{\alpha  \beta } g^{\mu  \nu } g^{\rho  \sigma }-g^{\alpha  \beta } \left(g^{\mu  \rho } g^{\nu  \sigma }+g^{\mu  \sigma } g^{\nu  \rho }\right)
		\\&
		C^{\mu\nu\rho\sigma\alpha\beta} \equiv \frac{1}{2} g^{\mu  \nu } \left(g^{\alpha  \rho } g^{\beta  \sigma }+g^{\alpha  \sigma } g^{\beta  \rho }\right)+\frac{1}{2} g^{\rho  \sigma } \left(g^{\alpha  \mu } g^{\beta  \nu }+g^{\alpha  \nu } g^{\beta  \mu }\right)
		\\&
		D^{\mu\nu\rho\sigma\alpha\beta}\equiv \frac{1}{2} \left(g^{\alpha  \sigma } g^{\beta  \mu } g^{\nu  \rho }+g^{\alpha  \rho } g^{\beta  \mu } g^{\nu  \sigma }+g^{\alpha  \sigma } g^{\beta  \nu } g^{\mu  \rho }+g^{\alpha  \rho } g^{\beta  \nu } g^{\mu  \sigma }\right)+\\&
		\qquad\qquad \qquad 
		\frac{1}{4} \left(g^{\alpha  \mu } g^{\beta  \sigma } g^{\nu  \rho }+g^{\alpha  \mu } g^{\beta  \rho } g^{\nu  \sigma }+g^{\alpha  \nu } g^{\beta  \sigma } g^{\mu  \rho }+g^{\alpha  \nu } g^{\beta  \rho } g^{\mu  \sigma }\right)
		\\&
		G^{\alpha \beta \gamma}\equiv \gamma^{\alpha }\gamma^{\beta }\gamma^{\gamma }-\gamma^{\beta 	}\gamma^{\alpha }\gamma^{\gamma }+\gamma^{\gamma }\gamma^{\alpha }\gamma^{\beta }-\gamma ^{\gamma }\gamma^{\beta }\gamma^{\alpha }
	\end{aligned}
\end{equation}
The vertices can then be written as

\begin{minipage}{0.28\textwidth}
	\includegraphics[width=0.93\linewidth]{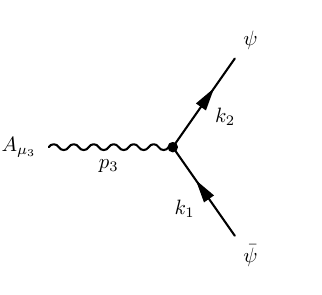}
\end{minipage}%
\begin{minipage}{0.5\textwidth}
	\begin{equation}
		\begin{aligned}
			V^{\mu_3}_{A\bar{\psi}\psi}=-i g \gamma^{\mu_3} \gamma_5  \nn
		\end{aligned}
	\end{equation}
\end{minipage}

\begin{minipage}{0.3\textwidth}
	\includegraphics[width=0.93\linewidth]{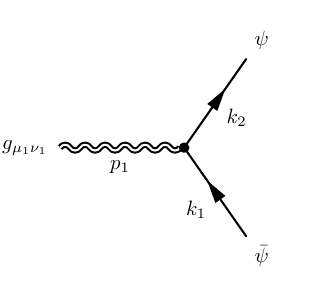}
\end{minipage}%
\begin{minipage}{0.6\textwidth}
	\begin{equation}
		\begin{aligned}
			V^{\mu_1 \nu_1}_{g \bar{\psi}\psi}=\frac{i}{4}A^{\mu_1 \nu_1 \rho \sigma}(k_1+k_2)_\rho \gamma_\sigma  \nn
		\end{aligned}
	\end{equation}
	\normalsize
\end{minipage}

\begin{minipage}{0.25\textwidth}
	\includegraphics[scale=0.87]{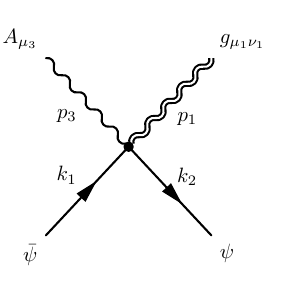}
\end{minipage}%
\begin{minipage}{0.68\textwidth}
	\begin{equation}
		\begin{aligned}
			V^{\mu_1 \nu_1 \mu_3}_{gA\bar{\psi}\psi}=-\frac{i g}{2}A^{\mu_1 \nu_1 \mu_3 \rho} \gamma_\rho \gamma_5 \nn
		\end{aligned}
	\end{equation}
	\normalsize
\end{minipage}

\begin{minipage}{0.35\textwidth}
	\includegraphics[scale=0.85]{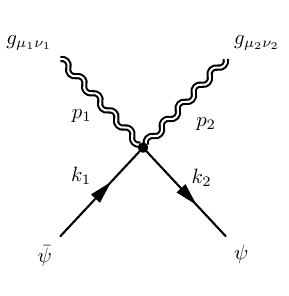}
\end{minipage}%
\begin{minipage}{0.55\textwidth}
	\begin{equation}
		\begin{aligned}
			&V^{\mu_1\nu_1\mu_2\nu_2}_{gg\bar{\psi}\psi}=\\&\quad -\frac{i }{8}\left(B^{\mu_1\nu_1\mu_2\nu_2 \alpha \beta}-C^{\mu_1\nu_1\mu_2\nu_2 \alpha \beta}+D^{\mu_1\nu_1\mu_2\nu_2 \alpha \beta}\right)\gamma_\alpha(k_1+k_2)_\beta\\&\quad 
			+\frac{i}{128}G^{\alpha \beta \gamma } A^{\mu_1\nu_1  \gamma \rho}p_2^\sigma \,\, \times \\&\qquad
			\left(g^{\alpha  \mu_2} g^{\beta  \sigma } g^{\nu_2 \rho }+g^{\alpha  \nu_2} g^{\beta  \sigma } g^{\mu_2 \rho }-g^{\alpha  \sigma } g^{\beta  \nu_2} g^{\mu_2 \rho }-g^{\alpha  \sigma } g^{\beta  {\mu_2}} g^{\nu_2 \rho }\right) \nn
		\end{aligned}
	\end{equation}
\end{minipage}

\begin{minipage}{0.35\textwidth}
	\includegraphics[scale=0.87]{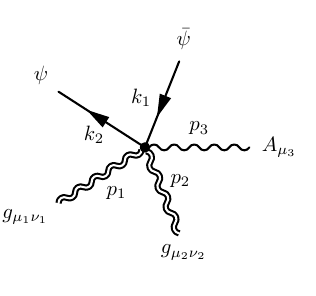}
\end{minipage}%
\begin{minipage}{0.6\textwidth}
	\begin{equation}
		\begin{aligned}
			&V^{\mu_1\nu_1\mu_2\nu_2\mu_3}_{ggA\bar{\psi}\psi}=
			\\&\qquad 
			-\frac{ig}{4} \left(B^{\mu_1\nu_1\mu_2\nu_2\mu_3 \lambda}-C^{\mu_1\nu_1\mu_2\nu_2\mu_3 \lambda}+D^{\mu_1\nu_1\mu_2\nu_2\mu_3 \lambda}\right) \gamma_{\lambda}\gamma_5\nn
		\end{aligned}
	\end{equation}
	\normalsize
\end{minipage}

\section{Definitions for the solution of the vacuum correlator} \label{app:defAij}
In this section, we provide a list of definitions for the functions introduced in eq$.$ \eqref{eq:pertresultff}. They are described as follows
\footnotesize
\begin{equation}
	\begin{aligned}
		A_{11}=&- \lambda \Bigg[2 p_1^{10}-p_1^8 (p_2^2+p_3^2)-2 p_1^6 (5 p_2^4-48 p_2^2 p_3^2+5 p_3^4)+4 p_1^4 (p_2^2+p_3^2) (4 p_2^4-23 p_2^2 p_3^2+4 p_3^4)\\&\qquad
		-8 p_1^2 (p_2^2-p_3^2)^2 (p_2^4+4 p_2^2 p_3^2+p_3^4)
		+(p_2^2-p_3^2)^4 (p_2^2+p_3^2)\Bigg]		\\
		A_{12}=&+2 p_2^2 \Bigg[p_3^2 (p_3^2-p_2^2)^5+p_1^{10} (38 p_3^2-12 p_2^2)+p_1^8 (18 p_2^4+41 p_2^2 p_3^2-121 p_3^4)
		-4 p_1^6 (3 p_2^6+46 p_2^4 p_3^2-38 p_2^2 p_3^4-26 p_3^6)\\&\qquad
		+p_1^4 (p_2-p_3) (p_2+p_3) (3 p_2^6+95 p_2^4 p_3^2+215 p_2^2 p_3^4+11 p_3^6)+14 p_1^2 p_3^2 (p_2^2-p_3^2)^3 (p_2^2+p_3^2)+3 p_1^{12}\Bigg]\\
		A_{13}=&+2 p_3^2 \Bigg[3 p_1^{12}+2 p_1^{10} (19 p_2^2-6 p_3^2)+p_1^8 (-121 p_2^4+41 p_2^2 p_3^2+18 p_3^4)+4 p_1^6 (26 p_2^6+38 p_2^4 p_3^2-46 p_2^2 p_3^4-3 p_3^6)\\&\qquad
		-14 p_1^2 p_2^2 (p_2^2-p_3^2)^3 (p_2^2+p_3^2)-p_1^4 (p_2-p_3) (p_2+p_3) (11 p_2^6+215 p_2^4 p_3^2+95 p_2^2 p_3^4+3 p_3^6)+p_2^2 (p_2^2-p_3^2)^5\Bigg]\\
		A_{14}=&-24 p_1^4 p_2^2 p_3^2 \Bigg[ (p_1^2-p_2^2)^3 (2 p_1^2+3 p_2^2)
		-3 p_3^4 (p_1^4+4 p_1^2 p_2^2-4 p_2^4)-3 p_3^2 (p_1^6-6 p_1^4 p_2^2+4 p_1^2 p_2^4+p_2^6)-3 p_3^8
		p_3^6 (7 p_1^2-3 p_2^2)\Bigg]\\
		A_{21}=&-\lambda \Bigg[2 p_3^6 (3 p_1^2+p_2^2)+4 p_1^2 p_3^4 (3 p_2^2-2 p_1^2)+(p_1^2-p_2^2)^4+2 p_3^2 (p_1-p_2) (p_1+p_2) (p_1^4+8 p_1^2 p_2^2+p_2^4)-p_3^8\Bigg]\\
		A_{22}=&-2 p_2^2 p_3^2 \Bigg[-17 p_1^8+p_1^6 (28 p_2^2+26 p_3^2)-4 p_1^4 (p_2^4+15 p_2^2 p_3^2)+(p_2^2-p_3^2)^4
		-2 p_1^2 (p_2^2-p_3^2)^2 (4 p_2^2+5 p_3^2)\Bigg] \\
		A_{23}=&+2 p_3^2 \Bigg[2 p_1^{10}-p_1^8 (p_2^2+6 p_3^2)+p_1^6 (-10 p_2^4+46 p_2^2 p_3^2+6 p_3^4)-2 p_1^2 (4 p_2^2+5 p_3^2) (p_2^3-p_2 p_3^2)^2+p_2^2 (p_2^2-p_3^2)^4
		\\&\qquad
		+2 p_1^4 (8 p_2^6-21 p_2^4 p_3^2-18 p_2^2 p_3^4-p_3^6)
		\Bigg]\\
		A_{24}=&-12 p_1^4 p_2^2 p_3^2 \Bigg[ p_3^4 (3 p_2^2-5 p_1^2)+(p_1^2-p_2^2)^3+p_3^2 (p_1^4+4 p_1^2 p_2^2-5 p_2^4)+3 p_3^6\Bigg] 
	\end{aligned}
\end{equation}
\normalsize
with the K\"{a}llen $\lambda$-function given by
\begin{equation}
	\lambda \equiv \lambda(p_1,p_2,p_3)=\left(p_1-p_2-p_3\right)\left(p_1+p_2-p_3\right)\left(p_1-p_2+p_3\right).\left(p_1+p_2+p_3\right)
	\label{Kallen}
\end{equation}

\section{Schouten identities}\label{app:schout}
When examining the gravitational anomaly corrections at finite density and temperature, we have introduced the rank-4 tensors $W_{i,j}^{\mu_1\nu_1\mu_2\nu_2}$. Such tensors are parity-odd and depend on the graviton momenta $p_1$ and $p_2$, as well as the momentum of the fermionic loop $l$.
There is a set of tensorial relations, known as Schouten identities, which can be used to reduce the expression of $W_{i,j}^{\mu_1\nu_1\mu_2\nu_2}$. These identities arise from the dimensional degeneracies of tensor structures, given that we are working in $d=4$.
In particular, in this case we can construct such identities starting from the following format
\begin{equation}
	0=\varepsilon^{[l \hspace{0.03cm} p_1 p_2  \mu_1} \delta^{\mu_2]\alpha}
\end{equation} 
Since we are antisymmetrizing over 5 indices and we are working in four dimensions, the result must vanish.
The index $\alpha$ can be contracted with a momentum, obtaining
\begin{equation}\label{eq:SImom}
	\begin{aligned}
		&0=\varepsilon^{[l \hspace{0.03cm} p_1 p_2  \mu_1} p_1^{\mu_2]}\\
		&0=\varepsilon^{[l \hspace{0.03cm} p_1 p_2  \mu_1} p_2^{\mu_2]}\\
		&0=\varepsilon^{[l \hspace{0.03cm} p_1 p_2  \mu_1} l^{\mu_2]}\\
	\end{aligned}
\end{equation}
or we can pick $\alpha=\{\nu_1, \nu_2\}$
\begin{equation}\label{eq:SIindex}
	\begin{aligned}
		&0=\varepsilon^{[l \hspace{0.03cm} p_1 p_2  \mu_1} \delta^{\mu_2]\nu_1}\\
		&0=\varepsilon^{[l \hspace{0.03cm} p_1 p_2  \mu_1} \delta^{\mu_2]\nu_2}
	\end{aligned}
\end{equation}
Note that we do not need to consider Schouten identities where both $\mu_1$ and $\nu_1$ are antisymmetrized since the energy-momentum tensor (and therefore $W_{i,j}^{\mu_1\nu_1\mu_2\nu_2}$) is symmetric under the exchange $\mu_1 \leftrightarrow \nu_1$. The same is true for the indices $\mu_2$ and $\nu_2$.\\
The identity \eqref{eq:SImom} and \eqref{eq:SIindex} relates tensors with rank less than four. Therefore, we need to complete them with the remaining indices. As an example we can pick the first identity in eq$.$ \eqref{eq:SIindex}, which is a rank-3 equation, and we multiply it with $p_1^{\nu_2}$, $p_2^{\nu_2}$ or $l^{\nu_2}$ ending up with three relations between rank-4 tensors. Proceeding in such way, we end up with 36 identities. Moreover, we can also get new identities from this 36 ones by exchanging $\mu_1 \leftrightarrow \nu_1$
and/or $\mu_2 \leftrightarrow \nu_2$. Therefore, the total number of Schouten identities is $36\times4=144$. All these identities can be used to simplify the structures appearing in the $W$  tensors.


\providecommand{\href}[2]{#2}\begingroup\raggedright\endgroup

\end{document}